\documentclass[12pt]{article}
\usepackage{epsf}
\usepackage{amsmath}

\usepackage{graphics}
\usepackage{cite}

\begin{titlepage}

\begin{document}

\hfill December 2021

\begin{center}
\large  
{\bf A Novel Viewpoint of Proton Decay}\\
\normalsize
\vspace{1.0in}

\vspace{1.0cm}
{\bf Paul H. Frampton\footnote{email:paul.h.frampton@gmail.com}\\}
\vspace{0.3cm}
{\it  Dipartimento di Fisica, Universit\`{a} del Salento \\
and  INFN Sezione di Lecce,  Via Arnesano 73100 Lecce, Italy\\}

\vspace{0.7in}

\begin{abstract}
\noindent
We update the standard model d=6 operators of Weinberg (1979) using a modified 
notation which accommodates the bilepton extension of the standard
model. This may lead to an enhancement of the proton lifetime by orders
of magnitude due to mixing first-family with third-family quarks.
By contrast, d=5 operators which can provide Majorana neutrino masses
retain the family structure of their counterparts in the standard model.
\end{abstract}

\end{center}

\end{titlepage}

\section{Introduction}

\noindent
Despite the fact that conservation of baryon number B is not a local symmetry, no example
of its violation has yet been established. The most readily observed example would be the
decay of the proton which was predicted by the earliest grand unified theories (GUTs) invented
in the 1970s. In 1974, minimal SU(5) \cite{GG} predicted a partial lifetime for the expected
dominant decay $p\rightarrow e^+\pi^0$ of $\tau_p \sim 10^{30.5}$ y.
A decade later, in 1984, experiment refuted this prediction
when the Irvine-Michigan-Brookhaven (IMB) collaboration
\cite{IMB} established a lower bound $\tau_p > 10^{32.5}$ y. The present limit for this decay
mode, provided by the Super-Kamiokande experiment\cite{SuperK}, is $\tau_p > 10^{34.5}$y.
As an upper limit on $\tau_p$, it is reasonable to use the first estimate of the proton lifetime by Sakharov
\cite{Sakharov} in 1967 based on gravitational interactions. It was $\tau_p \sim 10^{50}$y,
a lifetime beyond the reach of any terrestrial experiment using present technology.

\bigskip

\noindent
Before introducing our novel viewpoint, we briefly review proton decay in the minimal 
SU(5) GUT. This is based on the standard model and the only matter present from
the Fermi scale up to the GUT scale are the three quark-lepton families, treated
sequentially throughout this large energy range. The renormalisation group equations
(RGEs) suggested a value $M_{GUT}\sim 10^{15}$ GeV and a proton lifetime
$\tau_p \sim 10^{30.5}$y. The economy and simplicity of SU(5) was such that
a majority of the particle theory community supported it for the decade 1974-84
until the IMB result, already mentioned, slapped it down. Our novel viewpoint
will offer two related reasons for why SU(5) underestimated the proton lifetime
by at least four orders of magnitude.

\bigskip

\noindent
A useful analysis of B-violating higher-dimensional operators \cite{Weinberg1979}
 provides a more general approach not tied to SU(5)
and we shall make use of it in the present article. SU(5) assumed matter
was confined to the three-family standard model. Later we shall assume that the
matter representations are enlarged at the TeV scale as in the bilepton model 
\cite{Frampton1992} which offers a solution to the family puzzle and in which
a $|Q|=|L|=2$ gauge boson whose existence is subject to present LHC searches.  
We shall discuss how this could avoid the too-fast proton decay predicted by SU(5).

\begin{itemize}
\item The third family is no longer sequential, already at the TeV scale, with the
first two, so families must be treated asymmetrically in a GUT.
\item The d=6 operators suggest an enhancement factor $E > 10^4$ in proton lifetime because
of quark mixing with the third family.
\end{itemize}

\section{Modified Notation} 

\noindent
We begin by reviewing the d=6 $|\Delta B| \neq 0$ operators in the standard model using
Weinberg's notation from \cite{Weinberg1979}. For quarks we use
\begin{equation}
q_{i \alpha a L} ~~~~~~ i=1,2~~~ SU(2)_L;  ~~~~~\alpha=1,2,3  ~~ (colour); a=1,2,3 ~~~ (families).\\
\end{equation}
\noindent
and
\begin{equation}
u_{\alpha a R}; ~~~ d_{\alpha a R}
\end{equation}
\noindent
while, for leptons, we employ
\begin{equation}
l_{iaL} ~~~~~~~ i=1,2 ~~~ (SU(2)_L); ~~~~~~ a=1,2,3 ~~~ (families)
\end{equation}
\noindent
and
\begin{equation}
l_{a R}.
\end{equation}

\bigskip

\noindent
Then there are six d=6 B-violating operators, classified in \cite{Weinberg1979} as follows\\
\begin{eqnarray}
{\cal O}_{abcd}^{(1)} & = & (\bar{d}^c_{\alpha a R} u_{\beta b R})(\bar{q}^c_{i \gamma c L}l_{jdL}) \epsilon_{\alpha\beta\gamma} \epsilon_{ij}, \nonumber \\
{\cal O}_{abcd}^{(2)} & = & (\bar{q}^c_{i\alpha a L}q_{j \beta b L})(\bar{u}^c_{\gamma c R}l_{dR}) \epsilon_{\alpha\beta\gamma} \epsilon_{ij},\nonumber \\
{\cal O}_{abcd}^{(3)} & = & (\bar{q}^c_{i \alpha a L} q_{j \beta b L}) (\bar{q}^c_{k \gamma c L} l_{ i d L}) \epsilon_{\alpha\beta\gamma} \epsilon_{ij} \epsilon_{kl},\nonumber \\
{\cal O}_{abcd}^{(4)} & = & (\bar{q}^c_{i \alpha a L} q_{j \beta b L}) (\bar{q}^c_{k \gamma c L} l_{i d L})\epsilon_{\alpha\beta\gamma} (\underline{\tau} \epsilon)_{ij} . 
(\underline{\tau} \epsilon)_{kl}  ,\nonumber \\
{\cal O}_{abcd}^{(5)} & = & (\bar{d}^c_{\alpha a R} u_{\beta b R}) (\bar{u}^c_{\gamma c R} l_{dR}) \epsilon_{\alpha\beta\gamma} ,\nonumber \\
{\cal O}_{abcd}^{(6)} & = & (\bar{u}^c_{\alpha a R} u_{\beta b R}) (\bar{d}^c_{\gamma c R} l_{dR}) \epsilon_{\alpha\beta\gamma} . \nonumber \\
\label{d6}
\end{eqnarray}

\bigskip

\noindent
The well-known selection rule that $(B-L)$ must be conserved follows from Eq. (\ref{d6}), as do several other constraints..

\bigskip

\noindent
To study  the B-violating d=6 operators when the standard model is extended to the bilepton model\cite{Frampton1992}, we
introduce a modified notation as follows.

\noindent
A=1,2 (1st 2 families); ~~~ C=3 (3rd family); ~~~ I=1,2,3 ~~~ $SU(3)_L $

\noindent
Superscript I denotes triplet and subscript I denotes antitriplet.\\ \\

\bigskip

\noindent
\underline{Modified notation for quarks} \\

\noindent
For the first two families we write
\begin{equation}
q_{I \alpha A L}~~~~~ I=1,2,3~~~ SU(3)_L;  ~~~~~\alpha=1,2,3  ~~ (colour); A=1,2 ~~~ (families).
\end{equation}

\noindent
while for the third family we use

\begin{equation}
q^I_{ \alpha 3 L} ~~~~~~ I=1,2,3~~~ SU(3)_L;  ~~~~~ \alpha=1,2,3  ~~ (colour)
\end{equation}
Finally we add the $SU(3)_L$ singlets
 $u_{\alpha a R}; ~~~ d_{\alpha a R}$.\\ \\
\underline{Modified notation for leptons} \\

\noindent
The three families are here treated sequentially
\begin{equation}
L_{I a L} ~~~~~~~~~~~~ I=1,2,3~~~ SU(3)_L;  ~~~~~ a=1,2,3  ~~ (families).
\end{equation}

\section{d=6 Operators}

\noindent
Now we reconsider the six operators of Eq.(\ref{d6}).
Rewriting the ${\cal O}^{(n)}_{abcd}$ in the new notation reveals that the available d=6 operators
change significantly due to the new requirement that the operator be singlet under $SU(3)_L$.

\bigskip

\noindent
For example, consider trying to make an operator ${\cal O}^{(1)}_{abcd}$ using only the first
family. Such an operator ${\cal O}^{(1)}_{1111}$ does not exist because of the second parenthesis
where the lepton field is in a $\bar{3}$ of $SU(3)_L$ as is the quark field. To create an
overall $SU(3)_L$ singlet is possible only if the quark field is in a $3$ of $SU(3)_L$
which requires it to be in the third family. Thus, ${\cal O}^{(1)}_{1131}$ exists, although
${\cal O}^{(1)}_{1111}$ does not.

\bigskip

\noindent
A similar situation occurs, slightly differently, for ${\cal O}^{(2)}_{abcd}$. The lepton field
is in a $\bar{3}$ of $SU(2)_L$ and to make an overall $SU(3)_L$ singlet the first two
quark fields must be $\bar{3} \times \bar{3}$ which requires the second quark field to
be in the third family. Thus ${\cal O}^{(2)}_{1311}$ exists but ${\cal O}^{(2)}_{1111}$
does not.

\bigskip

\noindent
The two operators ${\cal O}^{(3,4)}_{abcd}$ are similar to each other. The lepton field
is in a $\bar{3}$ of $SU(3)_L$ while if we stay in the first family the three quark fields
are respectively in $(\bar{3}. 3, \bar{3})$ of $SU(3)_L$ and no singlet is possible. It
becomes possible only when either the first
or third quark belongs to the third family. Thus, ${\cal O}^{(3,4)}_{3111}$ and
${\cal O}^{(3,4)}_{1131}$ exist, but not ${\cal O}^{(3,4)}_{1111}$.

\bigskip

\noindent
Finally, the operators ${\cal O}^{(5,6)}_{abcd}$ do not exist as an $SU(3)_L$ singlet
for any choice of families because the three quark fields are singlets and the lepton
field is in a $\bar{3}$.

\bigskip

\noindent
To summarise, in the bilepton model none of the six ${\cal O}^{(n)}_{abcd}$ of Eq. (\ref{d6})
can form $SU(3)_L$ singlets when all four fermions are in the first or second family. With a quark
field in the third family, $n=1,2,3,4$
do exist, while $n=5,6$ are excluded for any family content.

\bigskip

\noindent
Weinberg also considered d=5 terms in the standard model:
\begin{equation}
f_{abmn}\bar{l}^c_{i a L} l_{jbL}\phi_k^{(m)}\phi_l^{(n)}\epsilon_{ik}\epsilon_{jl}
+f_{abmn}^{'}\bar{l}^c_{1 a L}l_{jbL}\phi_k^{(m)}\phi_l^{(n)}\epsilon_{ij}\epsilon_{kl}
\end{equation}
These terms violate lepton number and can contribute non-zero Majorana neutrino masses.

\bigskip

\noindent
Unlike the d=6 operators discussed above, the transition from the standard model
to the bilepton model does not make any essential difference in the family
structure for the d=5 operators and Majorana neutrino masses can arise in the
bilepton model just as 
discussed for the standard model in \cite{Weinberg1979}.

\bigskip

\section{Proton decay}

\noindent
The minimal bilepton model \cite{Frampton1992} does not, to our knowledge, fit into a GUT with a simple gauge group\footnote{No such model is known to us. We cannot prove that such a GUT is impossible.}. Nevertheless, it is  plausible to
assume that there exist unknown additional states which permit
unification as well as baryon-number-violating intermediaries. 

\bigskip

\noindent
In that case, proton decay will be generated by the d=6 operators discussed in the previous
section similarly to how the operators in Eqs. (\ref{d6}) generate proton decay in SU(5). In the
case of SU(5), as mentioned {\it ut supra}, the proton lifetime was predicted to be
some four orders of magnitude shorter than the present experimental lower limit. Several
theoretical attempts have been made to explain this discrepancy by some enhancement 
factor
in the lifetime, although no convincing argument was previously found.

\bigskip

\noindent
The bilepton model does provide a novel reason for a longer proton lifetime. The d=6 operators
require the appearance of a third-family quark. To mix with the first family quarks in the proton
this will lead, in the decay rate, to suppression by the square of the $(13)$-element of the
CKM quark mixing matrix. This element is of order $\sim \lambda^3$ where $\lambda$ is the 
sine of the Cabibbo
angle $\lambda = \sin \Theta_C \simeq 0.22$. This provides a
suppression of $\sim 10^{-2}$ in the matrix element, $\sim 10^{-4}$ in the decay rate
and hence an enhancement $E \sim 10^4$ in the proton lifetime.

\bigskip

\noindent
Since an explicit GUT is not at hand, further discussion necessarily involves an additional assumption.
The proton lifetime will be proportional to $E/M_{GUT}^4$ so that, in principle, $E$ can
be compensated by a change in $M_{GUT}$ but let us assume, for the moment,
that this does not happen
and that $M_{GUT}$ is close to its SU(5) value, $M_{GUT}\sim 10^{15}$ GeV. With such an assumption, the proton
lifetime would be predicted to be approximately $\tau_p \sim 10^{34.5}$Y which is at the lower
limit provided by Super-Kamiokande and within reach of Hyper-Kamiokande. Proton 
decay would provide important  information about far higher energies
than available in colliders.

\bigskip

\section{Discussion}

The bilepton model\cite{Frampton1992} gives a special r\^{o}le to the third family and
this could be reflected by the proton lifetime which is unexpectedly long from the viewpoint of
minimal SU(5). At least the assumption of three sequential families $3(10+\bar{5})$
between the Fermi and GUT scales is invalidated. In the above discussion, we saw
a novel reason that may enhance the proton lifetime is that the d=6 operators
require the costly mixing of first-family with third-family quarks.

\bigskip

\noindent
Such a crucial r\^{o}le of families is reminiscent of the
KM mechanism \cite{KM} for CP violation which is non-zero only if all the
three families participate as shown by the Jarlskog determinant \cite{Jarlskog}.

\bigskip

\noindent
The appearance of families first seemed, when the 
muon was discovered in 1936, to provide a major theoretical challenge but
now we
begin to understand how families can play an elegant
part in an emerging theory.

\newpage

\section{Acknowledgements}

\noindent
We are grateful for affiliation to the Department of Physics at the University of Salento in
Lecce, Italy. We thank F. Feruglio for a useful discussion.

\end{document}